\documentclass[reprint, one column, double space, times, 12pt, floatfix, superscriptaddress]{revtex4}

\usepackage{array}
\usepackage{graphicx}
\usepackage{appendix}
\usepackage{amsmath}
\usepackage{bm}
\usepackage{color}  
\usepackage{setspace}
\usepackage{tabularx}
\usepackage{natbib}

\begin{document}

\title{Phase-Field Modeling of Wetting and Balling Dynamics in Powder Bed Fusion Process}

\author{Lu Li}
\affiliation{Department of Mechanical Engineering, University of Connecticut, Storrs, CT
06269, USA}
\author{Ji-Qin Li}
\affiliation{Department of Mechanical Engineering, University of Connecticut, Storrs, CT
06269, USA}

\author{Tai-Hsi Fan } 
\thanks{Corresponding author. E-mail: thfan05@gmail.com}
\affiliation{Department of Mechanical
Engineering, University of Connecticut, Storrs, CT 06269, USA}

\begin{abstract}
\noindent In a powder bed fusion additive manufacturing (AM) process, the balling effect has a significant impact on the surface quality of the printing parts. Surface wetting helps the bonding between powder and substrate and the inter-particle fusion, whereas the balling effect forms large spheroidal beads around the laser beam and causes voids, discontinuities, and poor surface roughness during the printing process. To better understand the transient dynamics, a theoretical model with a simplified 2D configuration is developed to investigate the underlying fluid flow and heat transfer, phase transition, and interfacial instability along with the laser heating. We demonstrate that the degree of wetting and fast solidification counter-balance the balling effect, and the Rayleigh-Plateau flow instability plays an important role for cases with relatively low substrate wettability and high scanning rate. 
\end{abstract}
\date{\today}
\maketitle

\newpage
\section*{Introduction}
Surface quality is a great concern in making high-end 3D printing products. In a typical metal printing process a thin layer of powders is heated locally by a scanning laser or electron beam to selectively melt and fuse the powders or form a small region of the melt pool, and then followed by rapid solidification of the molten material. During each scan, a thin layer of powders is adhered to the substrate or to a previously solidified powder layer. A geometrically, structurally, or functionally complicated 3D configuration can be built by repeating this process, known as a layer-by-layer or solid freeform fabrication technique, which can be applied to a broad range of materials including alloys, ceramics, polymers, and composites. The fabrication technology has shown great promises in building light-weight structure from aerospace, automotive, to biomedical applications. For recent advances in additive manufacturing (AM) technology including materials, structures, processes, and the relevant multiscale physics see the comprehensive reviews~\cite{gu2012,yap2015,herzog2016,markl2016,malekipour2017,debroy2018}. The 3D printing technique may provide equivalent or superior microstructure, and thus the mechanical properties and performance, than conventionally cast and wrought materials~\cite{herzog2016}. However, the lack of consistent surface quality due to microstructural defects including pores, discontinuities, incomplete melting, process-induced microcracks, delamination, and balling-induced poor surface roughness hinder the advancement of the printing technology. Complicated interfacial phenomena in a metal AM process include multiple deforming interfaces, coalescence and change of morphology, conjugated heat and mass transfer, and multiscale phase transition dynamics. 

A critical concern regarding serious defects generated by the AM process is the balling phenomenon~\cite{malekipour2017,debroy2018}, which is the formation of large spheroidal beads and ripples from aggravated melting and solidification of metallic powders. Balling often appears around the scanning laser beam, resulting in a nonuniform adherence to the substrate or previously fused powder layer, including discontinuities or unfilled spots, which results in poor surface quality, poor interlayer bonding, and affects mechanical properties of the final parts. The balling phenomenon is primarily contributed by three factors: surface wettability, competition between spreading and solidification, and the Rayleigh-Plateau instability. In a few focused studies, Li et al.~\cite{li2012} observed the increase of balling tendency with higher oxygen content in the gas environment, higher scanning speed and powder layer thickness, and lower laser power, for both nickel and stainless steel powders. The process windows for tungsten and aluminum powders were observed by Wang et al.~\cite{wang2017} and Aversa et al.~\cite{aversa2018}, respectively. The former investigated the morphology and stability of a single scan track, whereas the latter suggested that balling may occur at either insufficient or excessive laser exposure, which is likely due to incomplete melting and denuded powders around the melt pool, respectively. This is consistent with the observations for iron-based powders~\cite{kruth2004,cherry2015,gunenthiram2017}. Surface oxidation reduces the wettability of the molten powders and may create a reversed (from negative to positive temperature coefficients) thermal Marangoni convection that further promotes balling~\cite{niu1999,rombouts2006}. Using high purity inert environment can prevent or reduce the content of surface oxides~\cite{das2003}. Agarwala et al.~\cite{agarwala1995} also suggested the addition of deoxidizer to improve wetting or applying the fluxing agent to enhance the fluidity of the molten metal during the fusion process. The competition between spreading and solidification plays an important role in transient fusion dynamics. Zhou et al.~\cite{zhou2015a} specifically compared the balling tendency and the resulting surface morphology under various laser parameters. They found that titanium and steel have faster spreading than solidification and are considered easy-to-process metals, whereas copper and tungsten have faster solidification time, implying a rapid solidification and arresting of the three-phase contact line, and thus a larger balling tendency. However, the new results found by Qiu et al.~\cite{qiu2020} on alumina ceramic powders indicated that severe balling could still happen even spreading is much faster than solidification. For materials with good wetting and spreading abilities, the Rayleigh-Plateau instability tends to break up the scan track and cause balling~\cite{rombouts2006,hunter2012}.  
Spattering under a very high laser power can also cause balling~\cite{gu2009}. The morphology of defects and discontinuities caused by balling can be reconstructed by 3D imaging using synchrotron radiation micro-CT~\cite{zhou2015b}. Many experimental observations have indicated that balling is primarily enhanced by three factors: 1) poor wetting ability, 2) higher tendency of solidification before spreading, and 3) onset of Rayleigh-Plateau instability. These factors are further complicated by local Marangoni convection due to surface oxidation and a large temperature gradient. 

Although a remelting procedure or tuning of the process window (including laser power, spot size, exposure time, hatch space, scanning speed and strategy) may help to avoid balling, a fundamental understanding and quantitative analysis of the interfacial phenomenon involved in balling are important for a better process design and control of the microstructure evolution during phase transition. Direct numerical simulation of selective laser melting process including fluid flow, phase transition, and heat transfer analysis has been successful using the volume of fluid (VOF) method, showing balling can be initiated by the Rayleigh-Plateau instability~\cite{khairallah2014,tang2018}. To better understand and quantify the influence of wettability, the competing dynamics of solidification and spreading, and interfacial instability to the balling effect, here we develop a theoretical model using an idealized 2D configuration with a single layer of powders on top of a substrate. The theoretical framework is developed based on the thermodynamically consistent phase-field method, which is broadly used for investigating the microstructure evolution in metallic systems such as growth kinetics and the formation of dendritic microstructure in material sciences~\cite{kobayashi1993,wheeler1993,warren1995,murray1995,karma1998,boettinger2002}. The phase-field method~\cite{cahn1958,cahn1961,penrose1990,wang1993,anderson1998,sekerka2011} has been broadened and applied to the mesoscale analysis of additive and pharmaceutical manufacturing processes that involve multiphase fluids, heat and mass transfer, thermal elasticity, phase transition, and three-phase contact line dynamics~\cite{li2018,li2019,fan2019,li2020a,li2020b}.

\section*{Theoretical Analysis}
In a simplified 2D configuration (Fig. 1) we consider a single layer of equal-sized metal powders aligned with the substrate. The powders are heated from the top by a laser beam with an assumed Gaussian irradiation heat flux. Further assumptions are made to facilitate the theoretical analysis: \romannumeral 1) evaporation kinetics and recoil pressure of the liquid metal are neglected at low to medium laser power, \romannumeral 2) the ambient argon gas is assumed ideal, \romannumeral 3) the gravity acceleration is neglected, \romannumeral 4) thermal elasticity is not considered, and \romannumeral 5) the latent heat, heat capacity, density of the metal powders are assumed constants, whereas the surface tension, dynamic viscosity, and thermal conductivity are temperature dependent. Starting from the definition of entropy functional, a concise mathematical framework that underlies the phase-field approach is provided here. More details about the derivations can be found in our recent work on the modeling of a laser brazing process~\cite{li2020b}.  

\begin{figure} [htbp!] \label{f1}
\centerline{\includegraphics[width=4.5in]{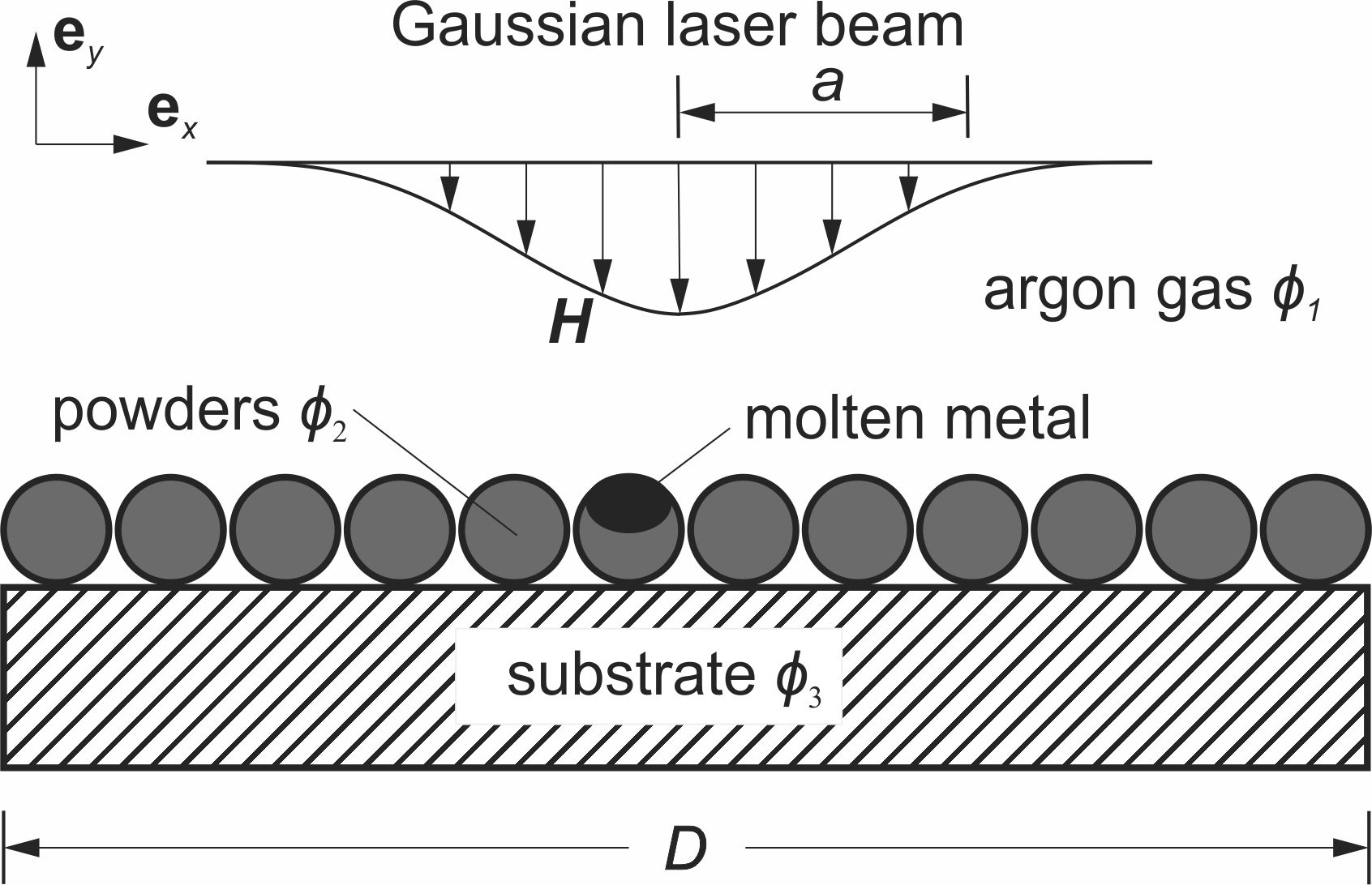}}
\caption{Schematic of selective laser melting of a horizontal layer of equal-sized metal powders placed on top of the substrate and is surrounded by argon gas. Phase field variables $\phi_1$, $\phi_2$, and $\phi_3$ indicate the volume fractions of argon gas, metal powders, and the substrate, respectively. $D$ represents the width of the computational domain, and the Gaussian beam is featured by its irradiation intensity $\textbf{\textit{H}}$ and a characteristic spot radius $a$.} 
\end{figure}

\subsection*{Entropy functional and free energy density}
As a starting point of thermodynamically consistent approach~\cite{penrose1990,wang1993,sekerka2011}, the entropy functional that describes the phase transition of a multi-component system can be expressed as
\begin{equation}
\begin{split}
    \ \mathcal{S}&=\int_\Omega \bigg[ s \left( e,\varphi,\phi_1,\phi_2,\phi_3\right)-\lambda \left(\sum_{i=1}^3 \phi_i-1 \right)\\ &~~~-\frac{1}{2}\xi_{\varphi}^2|\pmb{\nabla}\varphi|^2-\frac{1}{2}\sum_{i=1}^3\xi_i^2|\pmb{\nabla} \phi_i|^2 \bigg] dV~,
\end{split}
\end{equation}
where the overall material volume $\Omega$ indicates the computational domain shown in Fig. 1, including the substrate, metal powders, and the surrounding argon gas, $s$ is the local entropy density (per unit volume), $e$ is the internal energy, $\varphi\in[-1,1]$ is a non-conserved phase-field variable that describes solid-liquid phase transition of the powders (here $-1$ is for the liquid phase and $+1$ for the solid phase), $\phi_1$ to $\phi_3$ $\in[0,1]$ are material volume fractions for argon gas, metal powders, and the substrate, respectively, with the constraint of $\sum_{i=1}^3 \phi_i=1$, and $\lambda$ is the Lagrange multiplier. The gradient effects, $|\pmb{\nabla}\varphi|^2$ and $|\pmb{\nabla} \phi_i|^2$, along with their coefficients $\xi_{\varphi}$ and $\xi_{1 \sim 3}$ are associated with the interfacial energy, apparent thickness, and mobility. The transient evolution of the continuous phase field $\varphi$ and the volume fractions $\phi_1$ to $\phi_3$ are described by the phase-field equations, derived from the entropy transport equation and the above entropy functional. By requiring a positive entropy production rate, the time evolution of the non-conserved phase field $\varphi$ and conserved volume fraction $\phi_i$ can be formulated as
\begin{equation} \label{csl_b}
\frac{\partial \varphi}{\partial t} =M_{\varphi} \left( \frac{\partial s}{\partial \varphi} + \xi_{\varphi}^2 \nabla^2 \varphi \right)
\end{equation}
and
\begin{equation} \label{cfai_b}
\frac{\partial \phi_i}{\partial t} =-\pmb{\nabla} \cdot \left [M_i \pmb{\nabla} \left( \frac{\partial s}{\partial \phi_i} + \xi_i^2 \nabla^2 \phi_i \right) \right]~,
\end{equation}
respectively, where the assumed positive proportional constant $M_{\varphi}$ represents mobility at the solid-liquid interface of the melting powders, and $M_i$ represents mobilities of the interfaces between different compositions. 

The entropy in the partial derivatives in Eqs. (\ref{csl_b}) and (\ref{cfai_b}) are further associated with the free energy density (Appendix A). Here we consider the free energy that accommodates the enthalpy effect for the mixing of different components across the smooth interface and with an additional constraint from mass conservation~\cite{li2020b}, expressed as
\begin{equation} 
\begin{split}
f=&\sum_{i=1}^3\phi_i f_i+T\sum_{i=1}^3 h_i \phi_i^2(1-\phi_i)^2  \\
& +T\lambda \left( \sum_{i=1}^3 \phi_i -1 \right)~,
\end{split}
\end{equation}
where $f_i$ is the free energy of pure material, the 2nd term on the right assumes a double-well type mixing enthalpy with $h_i$ as the energy barriers, and the last term on the right takes the constraint in the entropy functional into account. The thermally driven phase transition dynamics, including melting and solidification, is described by the free energy density $f_{2}$ developed by Wang et al.~\cite{wang1993}, expressed as
\begin{equation} 
\ f_2=T \bigg[-\int_{T_m}^T \frac{e_2(T',\varphi)}{T'^2}dT'+\frac{1}{4}h_{\varphi}\left(1-\varphi^2\right)^2 \bigg]~,
\end{equation}
where $T_m$ is the equilibrium melting temperature of the pure metallic powders, $e_2=e_{2_s}+P(\varphi)L_a$~\cite{wang1993} is the corresponding internal energy to accommodate both solid and liquid phases of the metallic material using the interpolation function $P(\varphi)$ and the latent heat $L_a$. Here the polynomial interpolation function $P(\varphi)=1/2-(1/16)\left(3\varphi^5-10\varphi^3+15\varphi\right)$ has a range from $P(-1)=1$ to $P(1)=0$. $h_\varphi$ is the corresponding energy barrier across the solid and liquid phases. As a result, the phase field $\varphi$-equation that governs the solid-liquid phase transition~\cite{wang1993} can be formulated as
\begin{equation}
\begin{split}
\frac{\partial \varphi}{\partial t}  =M_{\varphi} &\bigg[\xi_{\varphi}^2\nabla^2\varphi +\phi_2 P'L_a\frac{T-T_m}{TT_m}\\ &~~~~+\phi_2 h_{\varphi} \left(\varphi-\varphi^3 \right)\bigg]~.
\end{split}
\end{equation}
The transient evolution of the phase field $\varphi$ is primarily determined by the thermally driving force based on the local temperature and the assumed temperature-independent latent heat, the balance of diffusive effect (the 1st term on the right) and double-well type phase separation (the 3rd term) for generating and evolving a smooth yet narrow interfacial profile. $P'$ is the $\varphi$-derivative of interpolation function $P$. The second phase-field equation that traces the volume fractions can be formulated as 
\begin{equation} \label{govi_b}
\begin{split}
\frac{\partial \phi_i}{\partial t} =\pmb{\nabla} \cdot \Bigg {\lbrace} M_i\pmb{\nabla}&\bigg[ 2h_i\phi_i(1-\phi_i)(1-2\phi_i)\\ &~~~~+\lambda-\xi_i^2\nabla^2\phi_i \bigg]\Bigg {\rbrace}
\end{split}
\end{equation}
for $i=1$ to $3$ in general. The first term on the right-hand side is originated from the double-well mixing enthalpy term, in which the energy barrier is adjustable numerically to prevent the mixing of different components in this case, and the Lagrange multiplier is resulting from the mass conservation constraint. The 4th-order term takes the long-ranged effect into account, which is obtained from the gradient terms that appeared in the entropy functional. In the above phase-field equations, the gradient coefficients $\xi_{\varphi}^2$ and $\xi_{i}^2$, and the energy barriers $h_{\varphi}$ and $h_i$ are associated with interfacial energy and the characteristic thickness of the interface, which will be explained in the following section. Note that to accommodate the fluid flow effect, hereafter we replace $\partial /\partial t$ by the substantial derivative $D /D t\equiv \partial /\partial t+\pmb v \cdot \pmb{\nabla}$ with $\pmb v$ indicating the velocity field.

\subsection*{Thermal energy equation} 
Following the thermodynamically-consistent formulation~\cite{penrose1990}, the differential energy equation associated with the entropy can be expressed as  
\begin{equation} \label{gove_b}
\frac{\partial e}{\partial t}=- \pmb{\nabla} \cdot  \left( M_e \pmb{\nabla} \frac{\partial s}{\partial e}\right)+\dot{Q}~,
\end{equation}
where the apparent mobility coefficient $M_e=T^2k_T(T)$, and $k_T$ is the temperature-dependent thermal conductivity to be determined by the material, phase, and temperature. The first term on the right-hand side reduces to the classical Fourier heat conduction effect, and the source term $\dot{Q}$ incorporates the radiation loss $\dot{Q}_{r}$ and laser irradiation $\dot{Q}_{ir}$ effects. The above internal energy has an additive form $e=\sum_{i=1}^3\phi_i e_i$, with $e_2=e_{2_s}+P(\varphi)L_a$. With further assumption that all specific heats $c_{p_i}$  remain constants, the energy equation (\ref{gove_b}) can be further written as
\begin{equation}
\begin{split}
\sum_{i=1}^3 \phi_i \rho_i c_{p_i} \frac{D T}{D t}  
& =\pmb{\nabla} \cdot (k_T \pmb{\nabla} T)+\dot{Q}_{r}+\dot{Q}_{ir}\\ &~~~~~~-\phi_2P' L_a \frac{D \varphi}{D t}~,
\end{split}
\end{equation}
where $\rho_i$ is the mass density. 

The exchange of thermal radiation energy between the material surface and ambient environment is simplified and expressed as
\begin{equation} 
\dot{Q}_{r}(\textbf{x}\in\partial\Omega)=-\frac{\epsilon \sigma_{\mbox{\tiny$B$}} (T^4-T_a^4)}{W}~,
\end{equation}
where $\epsilon$ is the emissivity of the metal surface with an apparent characteristic width $W$ in the phase-field model, $\sigma_{\mbox{\tiny$B$}}$ is the Stefan-Boltzmann constant, $T_a$ is the ambient temperature, $\alpha$ is the absorptivity of the metal material, assumed approximately the same as $\epsilon$. The gas absorption or participation is neglected. The irradiation of the laser beam on the powder surface is estimated as 
\begin{equation} 
\dot{Q}_{ir}(\textbf{x}\in\partial\Omega)=-\frac{\alpha \pmb H \cdot \pmb n}{W}~, 
\end{equation}
where $\pmb{n}$ is the surface normal computed by $\pmb{n}= \pmb{\nabla }\phi_1/|\pmb{\nabla} \phi_1|$, and $\pmb H$ is the intensity of an assumed 2D Gaussian laser beam, calculated by 
\begin{equation}
\pmb H=\frac{-\sqrt{2/\pi}\mathcal{Q}}{a} \textrm{exp} \left[ \frac{-2(x-x_0-U_at)^2}{a^2} \right] {\rm{\hat{\rm{}\textbf{e}}}_y}~,
\end{equation}
where $\mathcal{Q}$ is the laser power per unit width, $a$ is the characteristic spot radius, $x$ is the horizontal coordinate, $x_0$ is the initial laser focal point, and $U_a$ is the scanning speed of the laser beam traveling along the horizontal direction ($\hat{\textbf{e}}_x$ as shown in Fig. 1).

\subsection*{Interfacial dynamics}
In the phase-field approach, one has to determine a few characteristics of the smooth interface, including the interfacial energy, mobility, and apparent thickness. The interfacial energy, denoted by $\gamma$, is associated with the excess energy of the interface at equilibrium. As the interfacial thickness is much smaller than the feature size of the morphology during the phase transition, it is usually estimated by the 1D profile~\cite{cahn1958}, where the analytical solution of the equilibrium phase field is known. As a result, the interfacial energy at the solid-liquid interface is associated with the thickness, temperature, and the gradient coefficient as 
\begin{equation} \label{surfaceenergy}
\gamma_{\varphi} =\int_{-\infty}^\infty T_m\xi_{\varphi}^2|\pmb \nabla \varphi|^2dx=
 \frac{2\sqrt2}{3}\frac{\xi_{\varphi}^2}{W_{\varphi}}T_{m}~,
\end{equation}
where $x$ indicates the coordinate in an assumed unbounded 1D domain, $T_m$ is the reference temperature at the melting point of the material, and $W_{\varphi}$ is the characteristic thickness of interface correlated with the entropy gradient coefficient by setting $\xi_{\varphi}^2=h_{\varphi}W_{\varphi}^2$. With a further extension to a multi-component system, the energy barriers and gradient coefficients are related to the interfacial energies as
\begin{equation} \label{matrix}
\begin{bmatrix}
\xi_1^2 \\ \xi_2^2 \\ \xi_3^2 
\end{bmatrix}
=W^2
\begin{bmatrix}
h_1 \\ h_2\\ h_3
\end{bmatrix}
=\frac{3W}{\sqrt2 T_m}
\begin{bmatrix}
~1~&1~&-1~\\
~1~&-1~&1~\\
~-1~&1~&1~\\
\end{bmatrix}
\begin{bmatrix}
\gamma_{12} \\ \gamma_{13} \\ \gamma_{23}
\end{bmatrix},
\end{equation}
where the subscript 12, 13, and 23 indicate the gas-powder, gas-substrate, and power-substrate interfaces. To incorporate the thermal Marangoni effect, the interfacial energy between the nickel powder and the argon gas is calculated by a linear model $\gamma_{12}=\gamma_{12}^0-\beta_{\gamma}(T-T_m)$, with $\beta_{\gamma}$ as the temperature or Marangoni coefficient. The rest two interfacial energies $\gamma_{13}$ and $\gamma_{23}$ are treated as constants.The Lagrange multiplier $\lambda$ can be determined by combining Eq. (\ref{govi_b}), the constraint $\sum_{i=1}^3\phi_i=1$, and the assumption $M_1\xi_1^2= M_2\xi_2^2=M_3\xi_3^2$~\cite{boyer2006,boyer2011}, expressed as
\begin{equation}
\lambda=\frac{-1}{\sum_{i=1}^3 M_i}\left( \sum_{i=1}^3 2M_i h_i\phi_i(1-\phi_i)(1-2\phi_i)\right)~.
\end{equation}

We further assume that the mass density of nickel remains constant during the phase transition process, the molten nickel is a quasi-incompressible Newtonian fluid that satisfies the continuity equation: 
\begin{equation} 
\pmb{\nabla} \cdot \pmb v\simeq0~.
\end{equation}

Furthermore, the fluid dynamics involving the interfacial force can be described by the Naiver-Stokes-Korteweg momentum equation, written as
\begin{equation} 
\rho \frac{D \pmb v}{D t}=-\pmb \nabla \hat{p} +\pmb\nabla \cdot\left[\eta(\pmb\nabla \pmb v+\pmb\nabla \pmb v^T)\right]+\sum_{i=1}^3 \mu_i \pmb\nabla \phi_i +\textbf{F}_M ~,
\end{equation}
where $\hat{p}$ is a modified pressure, expressed as
\begin{equation}
\hat{p}=p-\sum_{i=1}^3 \left(T_m\xi_i^2\phi_i\nabla^2\phi_i\right)+f~,
\end{equation}
where $p$ is the hydrodynamic pressure, and the isotropic component of the interfacial force has been absorbed in the pressure gradient term for convenienc, $\eta$ is the temperature-dependent dynamic viscosity, $\mu_i=\partial f/ \partial \varphi-T \xi_i^2\nabla^2 \phi$ is the generalized chemical potential with $f$ indicating the bulk free energy, and $\textbf{F}_M$ is the body force due to the thermal Marangoni effect. Here we assume that the apparent Marangoni force $\textbf{F}_M$ is linearly proportional to the temperature gradient, within the region of interface selected by a scalar quantity $|\pmb \nabla \phi_1 |^2$ and written as
\begin{equation}
\textbf{F}_M\simeq\chi \pmb \nabla T |\pmb \nabla \phi_1 |^2, 
\end{equation}
where the coefficient $\chi$ is approximated by the increment of the surface tension through a simple control volume analysis around the interface, so that $\chi=-\beta_{\gamma} W$ with $\beta_{\gamma}$ obtained from the experimental Marangoni coefficient and $W$ as the characteristic thickness of the interface.

\subsection*{Material properties and parameters}

Because temperature variation influences the transport properties significantly, the nonlinear effect due to temperature-dependent properties is included in the computation. Here we summarize the material properties below for the case studies based on pure nickel powders and the argon gas environment. The thermal conductivity of nickel powders is determined by a smooth transition between solid and liquid phases using the $P$-function, expressed as
\begin{equation} 
k_{T_{Ni}}\simeq k_{Ni}^{(s)}\left[1-P(\varphi)\right]+k_{Ni}^{(\ell)}P(\varphi)~,
\end{equation}
where $k_{Ni}^{(s)}$ represents the solid-state thermal conductivity of pure nickel and is approximated by $k_{Ni}^{(s)}\simeq 50.06+0.022~T$~\cite{CRC1} in terms of dimensional values in MKS unit and degree Kelvin, and $k_{Ni}^{(\ell)}\simeq49.7~\textrm{W}/(\textrm{m} \cdot \textrm{K})$ is the liquid-state thermal conductivity of pure nickel. The dynamic viscosity of the molten nickel can be correlated with temperature~\cite{CRC1} as
\begin{equation}
\eta_{Ni}\simeq \eta_0+5.257\times 10^{-6}(T-T_m)~,
\end{equation}
where $T_m$ is the pure nickel's melting temperature, and $T$ is the absolute temperature field. The substrate material is assumed stainless steel with thermal conductivity~\cite{CRC1} estimated by:
\begin{equation}
k_{T_{Fe}}\simeq9.42+0.0143~T~.
\end{equation}
The thermal conductivity and dynamic viscosity of argon gas~\cite{eckhard2010} are approximated by
\begin{equation} 
k_{T_{Ar}}\simeq1.473\times10^{-2}+2.840\times10^{-5}~T
\end{equation}
and
\begin{equation} 
\eta_{Ar}\simeq1.885\times10^{-5}+3.362\times10^{-8}~T~,
\end{equation}
respectively. The density of argon gas is calculated by the ideal gas law. Table \ref{tab1} lists the assumed constant properties used for the case studies. Several characteristic lengths and model parameters are listed in Table \ref{tab2}. In summary, the fully coupled governing equations are in general applicable for 2D and 3D cases by solving the solid-liquid phase transition field $\varphi$, the volume fraction phase fields $\phi_{i (i=1,2,3)}$, temperature field $T$, and the velocity field $\pmb{v}$, along with the initial and boundary conditions.

\begin{table}[!htb]
\centering
\caption{Constant material properties.}
\label{tab1}
\begin{tabular}[t]{lr}
\hline
Parameters&Value\\
\hline
mass density:&$\textrm{kg}/\textrm{m}^3$\\
$\quad$ nickel $\rho_{Ni}$~\cite{CRC2}& $7810$\\
$\quad$ stainless steel $\rho_{Fe}$~\cite{CRC2}& $7874$\\
reference thermal conductivity $k_{T_0}$~\cite{CRC2} & $90.9$ $\textrm{W}/(\textrm{m} \cdot \textrm{K})$\\
specific heat:&$\textrm{J}/(\textrm{kg} \cdot \textrm{K})$\\
$\quad$ argon $c_{p_{Ar}}$~\cite{CRC2}& $520.3$\\
$\quad$ pure nickel $c_{p_{Ni}}$~\cite{CRC2}& $490.0$ \\
$\quad$ stainless steel $c_{p_{Fe}}$~\cite{CRC2}& $633.0$ \\
reference dynamic viscosity $\eta_0$~\cite{CRC2}& $5.01$ $\textrm{mPa} \cdot \textrm{s}$ \\
interfacial energy in between:&$\textrm{J}/\textrm{m}^2$\\
$\quad$ nickel and argon gas $\gamma_{12}^0$~\cite{CRC2} &1.838\\
$\quad$ nickel and stainless steel $\gamma_{23}$~\cite{CRC2} &2.385\\
$\quad$ stainless steel and argon gas $\gamma_{13}$~\cite{WH}&1.860\\
$\quad$ solid and liquid nickel $\gamma_{\varphi}$~\cite{jones2002} &0.347\\
Marangoni coefficient $\beta_{\gamma}$~\cite{CRC2} &$0.39$ $\textrm{mN}/(\textrm{m} \cdot \textrm{K})$ \\
melting temperature of nickel $T_m$~\cite{CRC1}&$1726$ K\\
latent heat of fusion of nickel $L_a$~\cite{CRC1} &$2.32 \times 10^6$ $\textrm{J}/\textrm{m}^3$\\
emissivity: \\
$\quad$ nickel powder $\epsilon_{Ni}$~\cite{MR}&$0.34$ \\
$\quad$ stainless steel $\epsilon_{Fe}$~\cite{MR}&$0.40$ \\
\hline
\end{tabular}
\begin{tabular}[t]{lr}
\renewcommand{\arraystretch}{1.2}
\noindent
\end{tabular}
\end{table}

\begin{table}[!htb]
\centering
\caption{Model parameters.}
\label{tab2}
\begin{tabular}[t]{lr}
\hline
Parameters&Value\\
\hline
characteristic length $L$& $200$ $\rm{\mu}m$\\
domain size $D=2 \pi L $&$\sim1200$ $\rm{\mu}m$\\
interfacial thickness for $\varphi$-field $W_{\varphi}$&$8$ $\rm{\mu}m$\\
interfacial thickness for $\phi_i$-field $W$&$4$ $\rm{\mu}m$\\
solid-liquid energy barrier $h_{\varphi}$&$18.5$ $\textrm{J}/(\textrm{m}^3\cdot\textrm{K})$\\
energy barrier for powders $h_2$&$91.7$ $\textrm{J}/(\textrm{m}^3\cdot\textrm{K})$\\
characteristic velocity $U$&0.73 $\textrm{m}/\textrm{s}$\\
interfacial mobility for $\varphi$-field  $M_{\varphi}$&32.1 $\textrm{m} \cdot \textrm{s} \cdot \textrm{K}/\textrm{kg}$ \\
mobility of nickel $M_2$&$0.259$ $\textrm{m}^3 \cdot \rm{\mu}\textrm{s} \cdot \textrm{K}/\textrm{kg}$ \\
2D power of laser beam $\mathcal{Q}$&$2.1\times10^5$ $\textrm{W}/\textrm{m}$\\
spot size of laser beam $a$&$100$ $\rm{\mu}\textrm{m}$\\
scanning speed of laser beam $U_a$&$0.1$ $\textrm{m}/\textrm{s}$\\
char. temperature difference $\Delta T$&$500$ K\\
\hline
\end{tabular}
\end{table}

Using the material properties and chosen model parameters above, a few characteristic time scales can be drawn: the thermal diffusion time scale $\tau_{\mbox{\tiny$T$}}=L^2 \rho_{0} c_{p_0}/k_{T_0}=1.68\times 10^{-3}~s$, convective time scale $\tau_c=L/U=2.74\times 10^{-4}~s$, solid-liquid phase transition time scale $\tau_{\varphi}=1/h_{\varphi} M_{\varphi}=1.68\times 10^{-3}~s$, time scale for wetting dynamics $\tau_\textrm{wet}=L^2/h_2 M_2=1.68\times 10^{-5}~s$, and the viscous diffusion time scale $\tau_\textrm{vis}=\rho_0 L^2/\eta_0=2.50\times 10^{-2}~s$~\cite{li2020b}.  Here we have applied characteristic length $L$, characteristic velocity $U$, characteristic overheating temperature $\Delta T$, phase transition time scale $\tau_{\varphi}$, and other reference material properties as provided in the tables. The reference parameters (with subscript 0) are based on powder material with density $\rho_0=\rho_{Ni}$ and specific heat $c_{p_0}=c_{p_{Ni}}$. The characteristic velocity $U$ is the scaled capillary velocity determined by $U=\beta\gamma_{12}/\eta_0$, with an adjustable factor $\beta=0.005$. For scaled formulation we suggest temperature to be scaled as
$\tilde{T}=(T-T_m)/\Delta T$, and pressure and stress to be scaled by the inertial effect $\rho_0 U^2$. We develop the computational solver based on Euler time integration and Fourier spectral discretization with $800\times800$ uniform mesh.

\section*{Results and Discussion}
\begin{figure*} [htbp!]
\centerline{\includegraphics[width=4.5in]{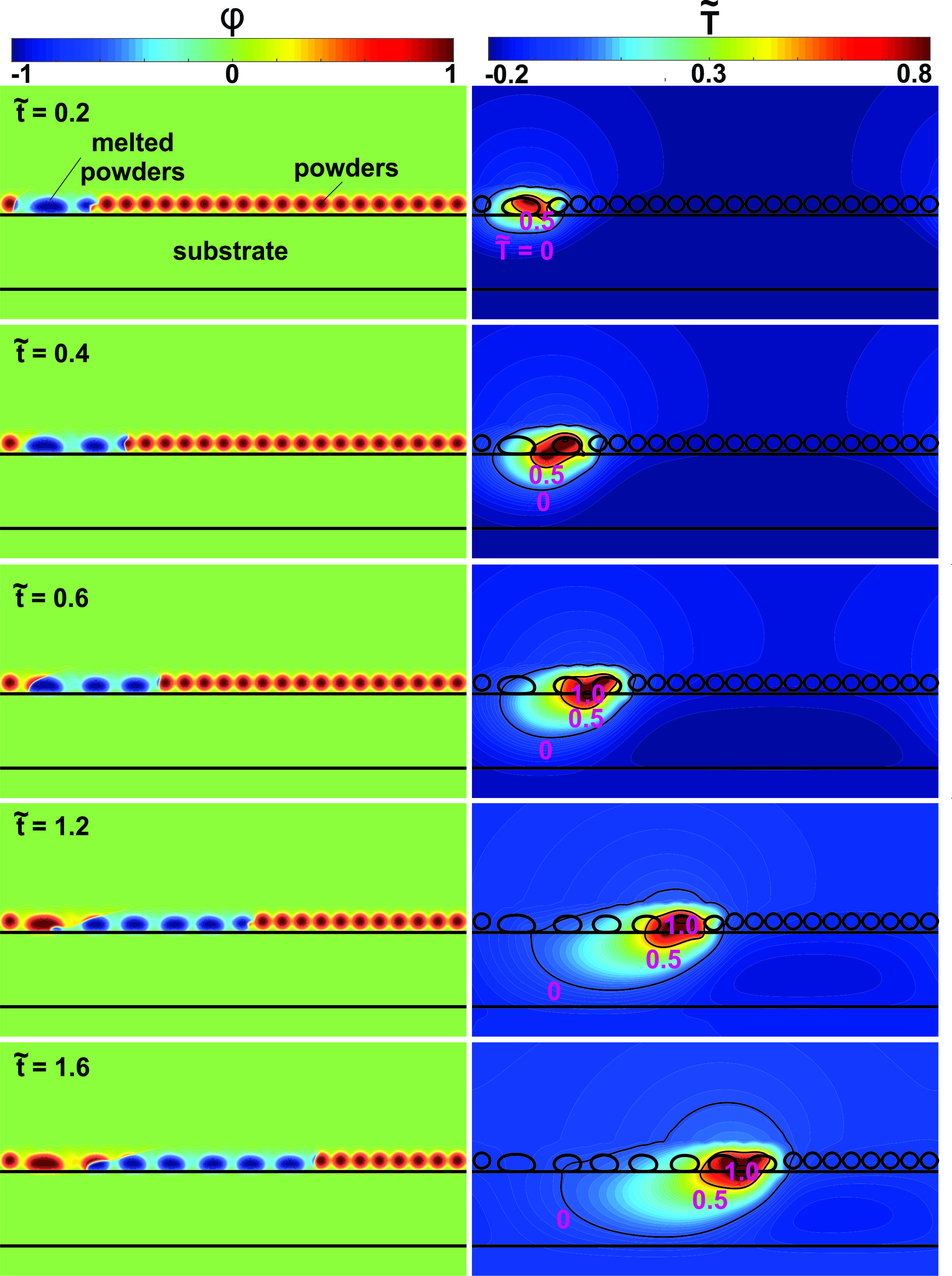}} 
\caption{Evolution of melting, wetting, and coalescence dynamics in terms of solid-liquid phase field $\varphi$ (left panels) and the corresponding temperature field on the right at five scaled time instants $\tilde{t}=0.2$, 0.4, 0.6, 1.2, and 1.6. The scaled temperature is defined as $\tilde{T}=(T-T_m)/\Delta T$, and the time is scaled by $\tau_{\varphi}=1.68\times 10^{-3}~s$. Three temperature contours $\tilde{T}=0$, 0.5, and 1.0 are provided for reference with the melting temperature at $\tilde{T}=0$. The laser parameters include: intensity $\mathcal{Q}=3.0\times 10^6~\rm{W}/ \rm{m}$, spot size $a=50~\mu m$, laser spot starting from $\tilde{x}=0.5$ and moving horizontally to the right with a scanning speed $U_a=0.5~\rm{m}/ \rm{s}$. } 
\end{figure*}

\begin{figure*} [htbp!]
\centerline{\includegraphics[width=\textwidth]{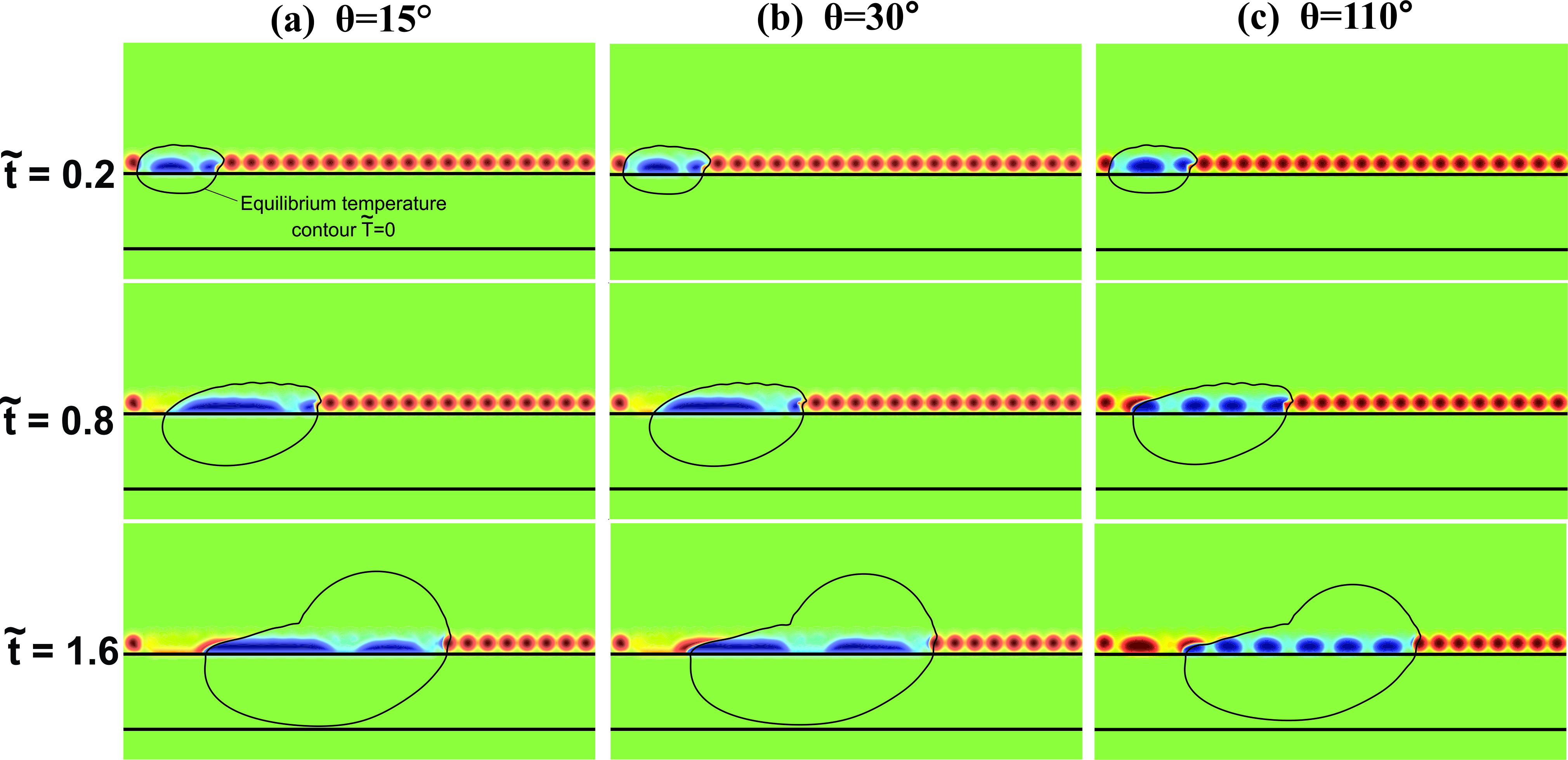}} 
\caption{Transient evolution of the solid-liquid interface at three scaled time instants $\tilde{t}=0.2$, 0.8, and 1.6 under three assumed contact angles: (a) $\theta=15^\circ$, (b) $\theta=30^\circ$, and (c) $\theta=110^\circ$. The temperature contour at $\tilde{T}=0$ is provided for reference.} 
\end{figure*}

Figure 2 shows the interplay between melting, wetting, and coalescence of a horizontal layer of metal powders. Melting and coalescence form larger droplets first, and then the poor wetting condition to the substrate with a contact angle $106.7^\circ$ further enhances the balling effect and creates a persistent pattern of separated balls. The process is followed by rapid solidification of the fused droplets without changing the pattern. A closer observation of the transient dynamics from the phase field shows that, initially ($\tilde{t}=0$) the whole system has uniform temperature $\tilde{T}=-0.2$, at $\tilde{t}=0.2$ four powders near the laser spot are fully melted, three of them are merged and the next one to the right splits due to the poor wetting condition and the cohesive force from the unmelted powders on the right. The overall profile of the melted powders has a blunt shape. Meanwhile, the ongoing and partially melted powder wets both the solid powder on the right and the substrate concurrently. The lower interfacial energy between the liquid and solid nickel powers essentially leads to a much better wetting condition for the laser melting process. The balling pattern is phenomenologically similar to Rayleigh-Plateau's classical work on the interfacial instability of a liquid column, however, here the characteristic distance between the gaps is roughly two times of the powder diameter, much smaller than the Rayleigh's criteria, which is about 4.5 times of the diameter of a 3D liquid column and without the influence of wetting and phase transition. Note that in practice the discontinuities in between droplets before and after solidification may create voids during the layer-by-layer AM process and should be avoided. Furthermore, the corresponding temperature distribution on the right panels shows the laser heating effect from Gaussian irradiation, phase transition, and thermal diffusion into the powders, substrate, and surrounding. The irradiation is absorbed by both powders and the substrate, whereas argon gas is assumed not participating in the radiation heat transfer. The system is thermally controlled and the melting front shown in the phase field coincides with the equilibrium temperature contour at $\tilde{T}=0$, which is consistent with our selection of time scale or interfacial mobility for the phase transition based on the thermal diffusion time scale. At dimensionless time $\tilde{t}=0.4$ and 0.6, metal powders are melted and merged, and soon started to solidify. The penetration depth of the thermal wave during the transient process can be used to characterize the heat affected zone. Finally, the morphology shows that the balling effect repeats itself through melting, coalescence, splitting, thermal relaxation, and solidification. The discontinuities that appeared in the balling pattern indicate a wavelength of around 60 to 65$\%$ of the wavelength obtained from a typical Raleigh instability, about 9 times of the radius of a thin liquid column. In addition to interfacial instability, the result suggests the importance of wetting and solidification to the formation of balling pattern. 

Figure 3 further demonstrates the balling effect under various wetting conditions. Under the same configuration and laser scanning speed, a better wetting condition with low contact angles (Fig. 3a) provides a flattened melt pool that coats the top surface of the substrate more uniformly than the high-contact-angle case. On the other hand, the number of discontinuities increases at a relatively poor wetting condition with a high contact angle (Fig. 3c). Note that the advancing and receding angles are not specified in this phase-field simulation, and the reference contact angle is defined based on the equilibrium condition. Overall, a good wetting condition inhibits the occurrence of balling and discontinuities. The morphology in general remains the same before and after solidification for each test case. The heat affected zone, featured by the melting temperature line within the substrate, appears quite similar for each case at a longer range. This is reasonable as balling is a very local phenomenon with a faster time scale and thus it has less influence on the temperature distribution in a larger heat affected zone.

\begin{figure*} [htbp!]
\centerline{\includegraphics[width=\textwidth]{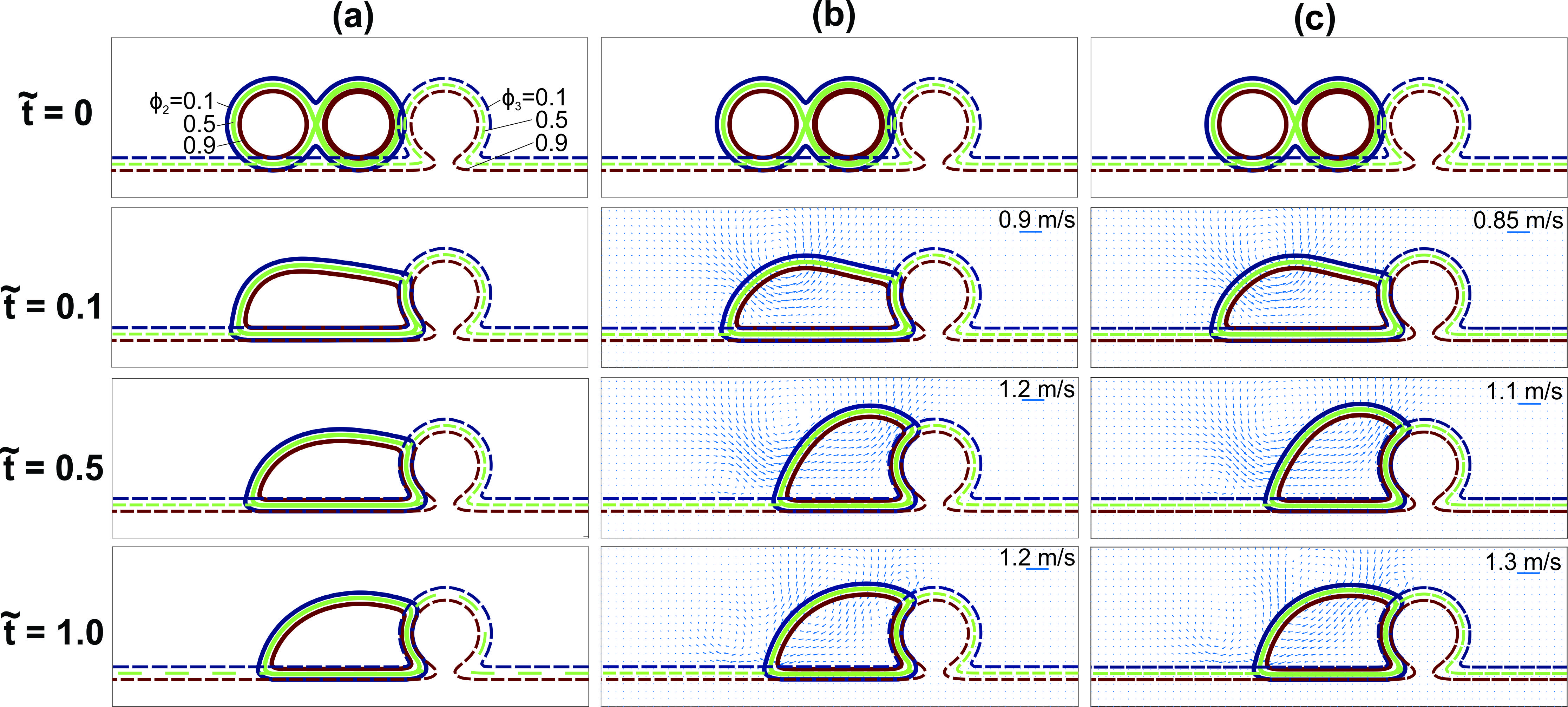}} 
\caption{Interfacial evolution showing wetting dynamics near a hypothetical solid powder afixed to the substrate at four sequential time instants for three cases: (a) without hydrodynamics, (b) including hydrodynamics but without Marangoni effect, and (c) including hydrodynamics and thermal Marangoni effect.  The profiles color coded by blue, green, and red are the contours for the volumetric fraction $\phi_2$=0.1, 0.5, and 0.9, respectively.} 
\end{figure*}

In Fig. 4 we apply the same test conditions as shown in Fig. 2 to have a closer look at the wetting dynamics. Note that for the case with a very low contact angle or a high wettability,  the lubrication approximation may be applicable to simplify the analysis, however, the full Navier-Stokes-Korteweg system is required in general and it is coupled with the phase transition dynamics in this study. Figure 4a shows the transient relaxation of the interfacial energy without the hydrodynamics. To simplify the study, first we consider two melted nickel droplets, described by the phase field $\phi_2$, placed on top of a solid substrate ($\phi_3$) and near a solid sphere made of stainless steel (also defined by $\phi_3$). Initially, the temperature is assumed uniform and above the melting temperature of pure nickel so that the heating and phase transition can be neglected. The equilibrium contact angle between the droplet and substrate is set to $\theta=60^\circ$. Figure 4b shows the results at the same time steps with fluid flow and convective effect taken into account. The characteristic velocity $U=0.92~m/s$ and reference Reynold number $Re=150$. The solid contour lines stand for the volume fraction of droplet at $\phi_2=0.1$, 0.5, and 0.9, whereas the dashed contour lines are for the volume fraction of stainless steel at $\phi_3=0.1$, 0.5, and 0.9. Apparently, fluid flow is driven by the moving three-phase contact line and the Laplace pressure around the free surface. The inertial and convective effects further accelerate the wetting, enhance heat transfer, and thus promote the phase transition process in terms of powder fusion and bonding efficiency to the substrate. However, due to the inertial effect, stronger oscillatory motion of the free surface along with few circulations around the droplet are introduced by the fluid flow inertial effect. One would expect more vigorous oscillation and even a spattering can happen near the melt pool under high laser power and long dwelling time, which is an important feature of the keyhole operation mode. Figure 4c includes the thermal Marangoni effect in the interfacial dynamics, and the result shows that in the test case Marangoni effect has a minor influence on the local velocity field. The result is reasonable since the characteristic Marangoni force is much smaller than the inertial force, i.e., $W \beta_{\gamma} \Delta T /(\rho_0 U^2 L^2)\sim 0.004$. The Marangoni effect is expected to be higher at high power laser melting process.

\section*{Conclusion}
We develop a thermodynamically consistent phase-field theoretical model to describe the balling effect that appeared in a simplified powder bed fusion process, which involves phase transition, wetting dynamics, interfacial deformation and instability, droplet coalescence, and the resulting discontinuities after solidification. The computational framework relies on the entropy functional of the system in addition to the conservation principles. All governing equations are derived to ensure positive local entropy generation. The fully coupled transport equations are solved by the spectral method. The result demonstrates that wetting condition and interfacial instability play major roles in the formation of discontinuities along the laser scanning track. A better wetting to the substrate reduces the degree of balling and the chance of discontinuity. At low to medium power laser heating, the solidification, inertial, and Marangoni convective effect around the micro melt pool are considered secondary effects to the balling pattern.   
\\\\
\noindent\textbf{Acknowledgments}~The authors acknowledge
the financial support of this research from the National Science Foundation (CBET 1930906).
\appendix
\label{appendix}
\section*{Derivative of internal energy and free energy} \label{Appendix A}
The total derivative of internal energy $e(s,\varphi,\phi_1,\phi_2,\phi_3)$ is expressed as 
\begin{equation} 
de=Tds+\frac{\partial e}{\partial \varphi} d\varphi+\sum_{i=1}^3 \frac{\partial e}{\partial \phi_i} d\phi_i~,
\end{equation}
and thus
\begin{equation}
ds=\frac{1}{T}de-\frac{1}{T} \frac{\partial e}{\partial \varphi} d\varphi-\frac{1}{T}\sum_{i=1}^3 \frac{\partial e}{\partial \phi_i} d\phi_i ~.
\end{equation}
By comparing the partial derivatives of entropy $s(e,\varphi,\phi_1,\phi_2,\phi_3)$ with the above expression, one can establish the following relations:
\begin{equation} 
\frac{\partial s}{\partial \varphi} \bigg)_{e,\phi_1,\phi_2,\phi_3}=-\frac{1}{T}\frac{\partial e}{\partial \varphi}\bigg)_{s,\phi_1,\phi_2,\phi_3}
\end{equation}
and
\begin{equation} 
\frac{\partial s}{\partial \phi_i} \bigg)_{e,\varphi,\phi_{j(j \neq i)}}=-\frac{1}{T}\frac{\partial e}{\partial \phi_i}\bigg)_{s,\varphi,\phi_{j(j \neq i)}}
\end{equation}
for $i=1$ to 3. Moreover, since the Helmholtz free energy density is introduced as $f(T, \varphi, \phi_1, \phi_2, \phi_3)=e-Ts$, the total derivative of the free energy is
\begin{equation} 
df=d(e-Ts)=-sdT+\frac{\partial e}{\partial \varphi} d\varphi+\sum_{i=1}^3 \frac{\partial e}{\partial \phi_i} d\phi_i~, 
\end{equation}
and therefore,
\begin{equation} 
\frac{\partial e}{\partial \varphi}  \bigg)_{s,\phi_1,\phi_2,\phi_3}=\frac{\partial f}{\partial \varphi}  \bigg)_{T,\phi_1,\phi_2,\phi_3}
\end{equation}
and
\begin{equation} 
\frac{\partial e}{\partial \phi_i} \bigg)_{s,\varphi,\phi_{j(j \neq i)}}=\frac{\partial f}{\partial \phi_i} \bigg)_{T,\varphi,\phi_{j(j \neq i)}}
\end{equation}
for $i=1$ to 3. Finally, 
\begin{equation} 
\frac{\partial s}{\partial \varphi} \bigg)_{e,\phi_1,\phi_2,\phi_3}=-\frac{1}{T}\frac{\partial f}{\partial \varphi}  \bigg)_{T,\phi_1,\phi_2,\phi_3}
\end{equation}
and
\begin{equation}
\frac{\partial s}{\partial \phi_i} \bigg)_{e,\varphi,\phi_{j(j \neq i)}}=-\frac{1}{T}\frac{\partial f}{\partial \phi_i} \bigg)_{T,\varphi,\phi_{j(j \neq i)}}
\end{equation}
for $i=1$ to 3.

\newpage
\bibliographystyle{plain}

\begin{thebibliography}{}
\begin{singlespace}
\bibitem{gu2012} D.D. Gu, W. Meiners, K. Wissenbach, R. Poprawe, Laser additive manufacturing of metallic components: materials, processes and mechanisms, \textit{Int. Mater. Rev.} \textbf{57}(3), 133-164, 2012.

\bibitem{yap2015} C.Y.~Yap, C.K.~Chua, Z.L.~Dong, Z.H.~Liu, D.Q.~Zhang, L.E.~Loh, S.L.~Sing, Review of selective laser melting: materials and applications, \textit{Appl. Phys. Rev.} \textbf{2}(4), 041101, 2015.

\bibitem{herzog2016} D.~Herzog, V.~Seyda, E.~Wycisk, C.~Emmelmann, Additive manufacturing of metals, \textit{Acta Mater.} \textbf{117}, 371-392, 2016.

\bibitem{markl2016} M.~Markl, C.~K\"orner, Multi-scale modeling of powder-bed-based additive manufacturing, \textit{Annu. Rev. Mater. Res.} \textbf{46}, 1-34, 2016.

\bibitem{malekipour2017} E.~Malekipour, H. El-Mounayri, Common defects and contributing parameters in powder bed fusion AM process and their classfication for online monitoring and control: a review, \textit{Int. J. Adv. Manuf. Technol.} \textbf{95}, 527-550, 2017.

\bibitem{debroy2018} T.~DebRoy, H.L.~Wei, J.S.~Zuback, T.~Mukherjee, J.W.~Elmer, J.O.~Milewski, A.M.~Beese, A.~Wilson-Heid, A.~De, W.~Zhang, Additive manufacturing of metallic components - process, structure and properties, \textit{Prog. Mater Sci.} \textbf{92}, 112-224, 2018.

\bibitem{li2012} R.~Li, J.~Liu, Y.~Shi, L.~Wang, W.~Jiang, Balling behavior of stainless steel and nickel powder during selective laser melting process, \textit{Int. J. Adv. Manuf. Technol.} \textbf{59}, 1025-1035, 2012.

\bibitem{wang2017} D.~Wang, C.~Yu, X.~Zhou, J.~Ma, W.~Liu, Z.~Shen, Dense pure tungsten fabricated by selective laser melting, \textit{Appl. Sci.} \textbf{7}, 430, 2017.

\bibitem{aversa2018} A.~Aversa, M.~Moshiri, E.~Librera, M.~Hadi, G.~Marchese, D.~Manfredi, M.~Lorusso, F.~Calignano, S.~Biamino, M.~Lombardi, M.~Pavese, Single scan track analyses on aluminium based powders, \textit{J. Mater. Process. Technol.} \textbf{255}, 17-25, 2018.

\bibitem{kruth2004} J.P.~Kruth, L.~Froyen, J.V.~Vaerenbergh, P.~Mercelis, M.~Rombouts, B.~Lauwers, Selective laser melting of iron-based powder, \textit{J. Mater. Process. Technol.} \textbf{149}, 616-622, 2004.

\bibitem{cherry2015} J.A.~Cherry, H.M.~Davies, S.~Mehmood, N.P.~Lavery, S.G.R.~Brown, J.~Sienz, Investigation into the effect of process parameters on microstructural and physical properties of 316L stainless steel parts by selective laser melting, \textit{Int. J. Adv. Manuf. Technol.} \textbf{76}, 869-879, 2015.

\bibitem{gunenthiram2017} V.~Gunenthiram, P.~Peyre, M.~Schneider, M.~Dal, F.~Coste, R.~Fabbro, Analysis of laser-melt pool-powder bed interaction during the selective laser melting of a stainless steel, \textit{J. Laser Appl.} \textbf{29}(2), 022303, 2017.

\bibitem{niu1999} H.J.~Niu, I.T.H.~Chang, Instability of scan tracks of selective laser sintering of high speed steel powder, \textit{Scr. Mater.} \textbf{41}(11), 1229-1234, 1999.

\bibitem{rombouts2006} M.~Rombouts, J.P.~Kruth, L.~Froyen, P.~Mercelis, Fundamentals of selective laser melting of alloyed steel powders, \textit{CIRP Ann.} \textbf{55}(1), 187-192, 2006.

\bibitem{hartnett2017} C.A.~Hartnett, I.~Seric, K.~Mahady, L.~Kondic, S.~Afkhami, J.D.~Fowlkes, P.D.~Rack, Exploiting the Marangoni effect to initiate instabilities and direct the assembly of liquid metal filaments, \textit{Langmuir} \textbf{33}, 8123-8128, 2017.

\bibitem{das2003} S.~Das, Physical aspects of process control in selective laser sintering of metals, \textit{Adv. Eng. Mater.} \textbf{5}(10), 701-711, 2003.

\bibitem{agarwala1995} M.~Agarwala, D.~Bourell, J.~Beaman, H.~Marcus, J.~Barlow, Direct selective laser sintering of metals, \textit{Rapid Prototyp. J.} \textbf{1}(1), 26-36, 1995.

\bibitem{zhou2015a} X.~Zhou, X.~Liu, D.~Zhang, Z.~Shen, W.~Liu, Balling phenomena in selective laser melted tungsten, \textit{J. Mater. Process. Technol.} \textbf{222}, 33-42, 2015.

\bibitem{qiu2020} Y.-D.~Qiu, J.-M.~Wu, A.-N.~Chen, P.~Chen, Y.~Yang, R.-Z.~Liu, G.~Chen, S.~Chen, Y.-S.~Shi, C.-H.~Li, Balling phenomenon and cracks in alumina creamics prepared by direct selective laser melting assisted with pressure treatment, \textit{Ceram. Int.}, in press, 2020.

\bibitem{hunter2012} R.~Mead-Hunter, A.J.C.~King, B.J.~Mullins, Plateau Rayleigh instability simulation, \textit{Langmuir} \textbf{28}, 6731-6735, 2012.

\bibitem{gu2009} D.~Gu, Y.~Shen, Balling phenomena in direct laser sintering of stainless steel powder: Metallurgical mechanisms and control methods, \textit{Mater. Des.} \textbf{30}, 2903-2910, 2009.

\bibitem{zhou2015b} X.~Zhou, D.~Wang, X.~Liu, D.~Zhang, S.~Qu, J.~Ma, G.~London, Z.~Shen, W.~Liu, 3D-imaging of selective laser melting defects in a Co-Cr-Mo alloy by synchrotron radiation micro-CT, \textit{Acta Mater.}, \textbf{98}, 1-16, 2015.

\bibitem{khairallah2014} S.A.~Khairallah, A.~Anderson, Mesoscopic simulation model of selective laser melting of stainless steel powder, \textit{J. Mater. Process. Technol.} \textbf{214}, 2627-2636, 2014.

\bibitem{tang2018} C.~Tang, J.L.~Tan, C.H.~Wong, A numerical investigation on the physical mechanisms of single track defects in selective laser melting, \textit{Int. J. Heat Mass Transfer} \textbf{126}, 957-968, 2018.

\bibitem{kobayashi1993} R. Kobayashi, Modeling and numerical simulations
of dendritic crystal growth, \textit{Physica D} \textbf{63}(3-4), 410-423, 1993.

\bibitem{wheeler1993} A.A~Wheeler, B.T.~Murray, R.J.~Schaefer, Computation of dendrites using a phase field model, \textit{Physica D} \textbf{66}, 243-262, 1993.

\bibitem{warren1995} J.A.~Warren, W.J.~Boettinger, Prediction of dendritic growth and microsegregation patterns in a binary alloy using the phase-field method, \textit{Acta Metall. Mater.} \textbf{43}(2), 689-703, 1995.

\bibitem{murray1995} B.T.~Murray, A.A~Wheeler, M.E.~Glicksman, Simulations of experimentally observed dendritic growth behavior using a phase-field model, \textit{J. Cryst. Growth} \textbf{154}, 386-400, 1995.

\bibitem{karma1998} A.~Karma, W.-J.~Rappel, Quantitative phase-field modeling of dendritic growth in two and three dimensions, \textit{Phys. Rev. E} \textbf{57}(4), 4323-4349, 1998.

\bibitem{boettinger2002} W.J.~Boettinger, J.A.~Warren, C.~Beckermann, A.~Karma, Phase-field simulation of solidification, \textit{Annu. Rev. Mater. Res.} \textbf{32}, 163-194, 2002.


\bibitem{cahn1958} J.W.~Cahn, J.E.~Hilliard, Free energy of a nonuniform system. I. interfacial free energy, \textit{J. Chem. Phys.} \textbf{28}(2), 258–267, 1958.

\bibitem{cahn1961} J.W.~Cahn, On spinodal decomposition, \textit{Acta Metall.} \textbf{9}(9), 795-801, 1961.

\bibitem{penrose1990}O.~Penrose, P.C.~Fife, Thermodynamically consistent models of phase-field type for the kinetics of phase transitions, \textit{Physica D} \textbf{43}(1), 44-62, 1990.

\bibitem{wang1993} S.-L.~Wang, R.F.~Sekerka, A.A.~Wheeler, B.T.~Murray, S.R.~Coriell, R.J.~Braun, G.B.~McFadden, Thermodynamically-consistent phase-field models for solidification, \textit{Physica D} \textbf{69}(1-2), 189-200, 1993.

\bibitem{anderson1998} D.M.~Anderson, G.B.~McFadden, A.A.~Wheeler, Diffuse-interface methods in fluid mechanics,~\textit{Annu. Rev.
Fluid Mech.} \textbf{30}(1), 139–165, 1998.

\bibitem{sekerka2011} R.F.~Sekerka, Irreversible thermodynamic basis of phase field models,~\textit{Philos. Mag.} \textbf{91}(1), 3-23, 2011.

\bibitem{li2018} J.-Q.~Li, T.-H.~Fan, T.~Taniguchi, B.~Zhang, Phase-field modeling on laser melting of a metallic powder,  \textit{Int. J. Heat Mass Transfer} \textbf{117}, 412-424, 2018. 

\bibitem{li2019} J.-Q.~Li, T.-H.~Fan, Phase-field modeling of metallic powder-substrate interaction in laser melting process, \textit{Int. J. Heat Mass Transfer} \textbf{133}, 872-884, 2019.

\bibitem{fan2019} T.-H. Fan, J.-Q. Li, B.~Minatovicz, E.~Soha, L.~Sun, S.~Patel, B.~Chaudhuri, R.~Bogner. Phase-field modeling of freeze concentration of protein solutions, \textit{Polymers} \textbf{11}(1), 10, 2019.

\bibitem{li2020a} J.-Q.~Li, T.-H.~Fan, Phase-field modeling of macroscopic freezing dynamics in a cylindrical vessel, \textit{Int. J. Heat Mass Transfer}, \textbf{156}, 119915, 2020.

\bibitem{li2020b} L.~Li, S. Li, B. Zhang, T.-H.~Fan, Phase-Field Modeling of Selective Laser Brazing of Diamond Grits, in review, 2020.

\bibitem{wu2019} Y.-C.~Wu, F.~Wang, M.~Selzer, B.~Nestler, Investigation of equilibrium droplet shapes on chemically striped patterned surface using phase-field method, \textit{Langmuir} \textbf{35}, 8500-8516, 2019.

\bibitem{xu2020} D.~Xu, Y.~Ba, J.-J.~Sun, X.-J.~Fu, A numerical study of micro-droplet spreading behaviors on wettability-confined tracks using a three-dimensionless phase-field lattice Boltzmann model, \textit{Langmuir} \textbf{36}, 340-353, 2020.

\bibitem{yeh2015} S.Y.~Yeh, C.W.~Lan, Adaptive phase-field modeling of anisotropic wetting with line tension at the triple junction, \textit{Langmuir} \textbf{31}, 9348-9355, 2015.

\bibitem{boyer2006} F.~Boyer, C.~Lapuerta, Study of a three component Cahn-Hilliard flow model, \textit{ESAIM: Math. Model. Numer. Anal.} \textbf{40}(4), 653-687, 2006.

\bibitem{boyer2011} F.~Boyer, S.~Minjeaud, Numerical schemes for a three component Cahn-Hilliard model, \textit{ESAIM: Math. Model. Numer. Anal.} \textbf{45}(4), 697-738, 2010.

\bibitem{CRC1} J.F.~Shackelford (Ed.) \textit{CRC Materials Science and Engineering Handbook}, CRC Press, Boca Raton, 4th edition, 2016.

\bibitem{eckhard2010} E.~Vogel, B.~J\"{a}ger, R.~Hellmann, E.~Bich, \textit{ Ab~initio} pair potential energy curve for the argon atom pair and thermophysical properties for the dilute argon gas. II. Thermophysical properties for low-density argon, \textit{Mol. Phys.} \textbf{108}(24), 3335-3352, 2010.

\bibitem{CRC2} W.M.~Haynes (Ed.) \textit{CRC Handbook of Chemistry and Physics}, CRC Press, Boca Raton, 2012.

\bibitem{WH} N.~Eustathopoulos, M.G.~Nicholas, B.~Drevet, \textit{Wettability at High Temperatures}, Elsevier, Amsterdam, 1999.

\bibitem{jones2002} H.~Jones, The solid-liquid interfacial energy of metals: calculations versus measurements, \textit{Mater. Lett.} \textbf{53}, 364-366, 2002.

\bibitem{MR} W.F.~Gale, T.C.~Totemeier(Eds), \textit{Smithells Metals Reference Book}, Elsevier, Amsterdam, 2004.


\end{singlespace}
\end{thebibliography}
{}
\end{document}